\documentclass[10pt,journal,compsoc]{IEEEtran}
\ifCLASSOPTIONcompsoc
\usepackage[nocompress]{cite}
\else
\usepackage{cite}
\fi
\usepackage{color}
\usepackage{graphicx}
\usepackage{amsmath}
\usepackage{amsfonts}
\usepackage{tabularx} % tabularx
\usepackage{booktabs} % \toprule, \midrule, \bottomrule
\usepackage{array}
\ifCLASSINFOpdf
 \DeclareGraphicsExtensions{.pdf,.jpeg,.png}
\else
\fi

\begin{document}
\title{Interest-Behaviour Multiplicative Network for Resource-limited Recommendation}

%\author{Qianliang Wu, Tong Zhang, Zhen Cui, Jian Yang}
%%\authornotemark[1]
%%\email{webmaster@marysville-ohio.com}
%
%\affiliation{%
%	\institution{PCA Lab, Key Lab of Intelligent Perception and Systems for High-Dimensional Information of Ministry of Education, and Jiangsu Key Lab of Image and Video Understanding for Social Security, 
%		\\School of Computer Science and Engineering, Nanjing University of Science and Technology, Nanjing, China}
%	%\email{{yun.wang, tong.zhang, zhen.cui, cyx, csjyang}@njust.edu.cn}
%	\city{\{wuqianliang, tong.zhang, zhen.cui, csjyang\}@njust.edu.cn}
%}

%\author{
%	\IEEEauthorblockN{Qianliang Wu, Tong Zhang, Zhen Cui, Jian Yang}
%	\IEEEauthorblockA{
%	PCA Lab, 
%	Key Lab of Intelligent Perception and Systems for High-Dimensional Information of Ministry of Education, \\
%	Jiangsu Key Lab of Image and Video Understanding for Social Security, \\
%	School of Computer Science and Engineering, \\
%	Nanjing University of Science and Technology, Nanjing, China\\
%	Email: \{wuqianliang, tong.zhang, zhen.cui, csjyang\}@njust.edu.cn}
%
%}

\author{Qianliang Wu,
		Tong Zhang,
		Zhen Cui,
		Jian Yang
	\IEEEcompsocitemizethanks{\IEEEcompsocthanksitem Q. Wu, T. Zhang, Z. Cui and J. Yang are with the Key Laboratory
		of Intelligent Perception and Systems for High-Dimensional Information of
		Ministry of Education, School of Computer Science and Engineering, Nanjing
		University of Science and Technology, Nanjing, 210094, China.
		Email: \{wuqianliang, tong.zhang, zhen.cui, csjyang\}@njust.edu.cn}% <-this % stops an unwanted space
	}

\IEEEtitleabstractindextext{%
	\begin{abstract}
Resource constraints, e.g. limited product inventory or financial strength, may affect consumers' choices or preferences in some recommendation tasks but are usually ignored in previous recommendation methods.  In this paper, we aim to mine the cue of user preferences in resource-limited recommendation tasks, for which purpose we specifically build a large used car transaction dataset possessing resource-limitation characteristics. Accordingly, we propose an interest-behavior multiplicative network to predict the user's future interaction based on dynamic connections between users and items. To describe the user-item connection dynamically, mutually-recursive recurrent neural networks (MRRNNs) are introduced to capture interactive long-term dependencies, and meantime effective representations of users and items are obtained. To further take the resource limitation into consideration, a resource-limited branch is built to specifically explore the influence of resource variation on user preferences. Finally, mutual information is introduced to measure the similarity between the user action and fused features to predict future interaction, where the fused features come from both MRRNNs and resource-limited branches. We test the performance on the built used car transaction dataset as well as the Tmall dataset, and the experimental results verify the effectiveness of our framework.
	\end{abstract}
}

\maketitle

\section{Introduction}
Nowadays, recommendation systems have become pivotal assist tools for users to filter information and locate their preferences on e-commercial websites. Based on browsing and transaction records of users during online shopping, recommendation systems aim to capture the interests of users accurately, and further recommend multiple products that users may be interested in. In this way, recommendation systems effectively promotes online shopping experience by facilitating users to find interesting products, and create opportunities for the e-commercial website to increase revenue. Because of these advantages, online shopping has become a habit and fashion of consumption.

Numerous algorithms have been proposed for user recommendation in previous literatures~\cite{Kang2018SelfAttentiveSR,He2017NeuralCF,7920291,7069201,8693783,6004833,rao2015collaborative,koren2009matrix,pazzani2007content,bai2017dltsr,van2013deep,oh2014personalized,wang2018content,Bai2019ALD,Yuan2019ASC,Wang2019ModelingIT}. Some of them~\cite{tomassini2007empirical,rao2015collaborative,koren2009matrix} characterize users' interests based on collected ratings of items, and further treating recommendation as matrix completion problem for mining the rating patterns. And some others~\cite{pazzani2007content,bai2017dltsr,van2013deep,oh2014personalized,wang2018content} employ deep neural networks to derive similarities between items and users to recommend new items. Specifically, considering the dynamics of online shopping processes, a part of works~\cite{Hidasi2015SessionbasedRW,Smirnova2017ContextualSM,Quadrana2017PersonalizingSR} employ recurrent neural networks (RNNs) to capture the user-item interest evolution in one period based on existing user-item interaction records.   

Great success has been achieved by the aforementioned methods, however, there are still various challenges existing in real-world applications. One crucial problem is the significant impact of resource limitation on user interests, which is always ignored in previous works. In general, the resource limitation exists in both products and users, e.g. inventory constraint and limited financial strength. For instance, after purchasing valuables, in one period, users' desire to buy other high-end products may decrease due to financial limitation. Also, consumers may turn to other suboptimal products when the most interesting ones are out of inventory. Therefore, it is non-trivial to consider the resource limitation in designing recommendation models, especially to appropriately pick out resource limitation related information and further model them with effective algorithms. However, there are multiple issues to be tackled in this process, including two crucial ones:
\begin{enumerate}
\item[(i)]  Suitable datasets. Existing recommendation datasets usually focus on recording user action and product attributes, but often ignoring the collection of resource constraint related information. This makes them hardly be employed for the investigation of resource-limited user recommendation tasks. 
\item[(ii)]  Effective inference on the intricate system. The complicated information of resource limitation, the dynamics of user-item interaction, and the similarity between each pair of user and product indicating the user preference, should be well represented and further modeled in the designed framework.
\end{enumerate}

Based on the analysis above, in this paper, we investigate the resource-limited user recommendation system, and propose an interest-behaviour multiplicative neural network (IMN-Net) to mine the cue of the user's preferences for products. First, we specifically construct a large recommendation dataset named `CheZhiBao' about used car transactions ('Cars'\footnote{https://www.cars.com/}) possessing resource-limitation characteristics. As a special recommendation task, used car recommendation based on `CheZhiBao' dataset well meets the resource-limited situation from two main aspects: (1) limited supply of products: obviously, the inventory of each product is not constant, and restricted by those used car owners; (2) deep pocket requirement: the high price of used cars makes it unrealistic for users, even for some companies, to frequently make orders just like buying daily products. Then, to model the system, an interest-behaviour multiplicative network is accordingly constructed. Considering dynamic user-item connections, we introduce the mutually-recursive recurrent neural networks (MRRNNs)  to capture interactive long-term dependencies of users and items, and meanwhile extract high-level features of them. To model the resource limitation, we build another resource-limited branch to specifically explore the influence of resource variation caused by user behaviour. Specifically, a RNN branch is employed to learn the representation from the variation of user and product status once a transaction happens. Finally, the features of both MRRNNs and resource-limited branch are fused, and further used to measure the similarity between user interest and the product. Specifically, here, mutual information is introduced for user-product similarity measurement in an unsupervised manner.  For the optimization of the entire framework, both supervised and unsupervised losses are calculated based on which the parameters are tuned through back-propagation. In our experiments, the performance is tested on the built `CheZhiBao'  dataset as well as the Tmall dataset, and the results verify the effectiveness of our proposed framework.       
  
To summarize, in this paper, our contributions are three-fold:
%\begin{itemize}
%\item {\verb|textbf{A new used car dataset}|}
\begin{enumerate}
	\item[(i)] \textbf{A new used car dataset}. We construct a new large-scale used car dataset based on the records of an online shopping platform\footnote{https://www.chezhibao.com/}. To the best of our knowledge, this is the first dataset for the used car recommendation task with resource-limitation characteristics.
	\item[(ii)] \textbf{Resource-limitation modeling}. We specifically investigate the resource-limitation problem which is always ignored in previous literatures. Accordingly, we propose the novel IMN-Net to model the intricate recommendation system by using a special resource-limited branch to capture the influence of resource variation on user interests.
	\item[(iii)] \textbf{Similarity measurement}. We introduce mutual information to measure the similarities between items and users' actions. Considering the informative representation of items and users, mutual information may be more suitable in the similarity measurement as it considers the associated distribution of the representation.
    \item[(iv)] \textbf{Effectiveness}. We test the proposed framework on our built 'CheZhiBao' as well as Tmall datasets. The experimental results verify the effectiveness of our proposed IMN-Net. 
\end{enumerate}

\section{Related work}
We aim to design an effective sequential learning system for the special resource-limited recommendation task. To the best of our knowledge, few works have been proposed to investigate this specific task especially based on online used car transaction. Instead, when it comes to used car related research or application, most previous literatures just focus on used car transaction system construction~\cite{Lee1997AUCNETEI,Lhe2009AssessmentOS}, market research~\cite{Emons2009TheMF}, and price evaluation~\cite{Kooreman2002ThreePA,Haan2010HasTI,Pudaruth2006PredictingTP}, etc. Therefore, we  mainly introduce existing user recommendation related algorithms from the technique perspective rather than application, including deep neural networks and mutual information (for similarity measurement).

In early stage of recommendation system, recommendation algorithms mainly include the Markov chain~\cite{Rendle2010FactorizingPM, Ruining2016FusingSM}, and matrix factorization \cite{He2017NeuralCF}. Factorization based approaches decompose user-item interaction matrix built from users' feedback to obtain low-rank embedding of users/items, then make matrix completion or subsequent predictions via the inner product of user and item embedding vectors. These kinds of methods did not consider the time order of interactions, therefore are not suitable for the sequential recommendation scenarios. 

Recently, deep neural network models, e.g. RNNs and CNNs,  have achieved significant success in contrast to conventional methods. RNNs, based on the original intention of its design, have become the most successful one in sequence recommendation~\cite{Hidasi2015SessionbasedRW}.
Several RNN variants~\cite{Zhu2017WhatTD,Quadrana2017PersonalizingSR,Smirnova2017ContextualSM,Yu2019AdaptiveUM,Hidasi2015SessionbasedRW} have been proposed to adapt to different application scenarios. On the other hand, considering the great success of CNN model in computer vision, there have been various CNN-based sequential recommendation models~\cite{Yuan2019ASC,Yan2019CosRec2C,Tang2018PersonalizedTS}  proposed to date. As attention-mechanism achieved promising performance in NLP~\cite{Vaswani2017AttentionIA,Bahdanau2014NeuralMT,Song2019MASSMS},  many attention-based sequential models~\cite{Ying2018SequentialRS,Yuan2020AttentionbasedCS,Xu2019GraphCS} have also been developed. In particular, the self-attention model~\cite{Kang2018SelfAttentiveSR} showed powerful ability in sequential recommendation system of daily consuming e-commerce. Most of them seek to learn effective representation based on a suboptimal prior or to capture similarities between interacted items in one period. However,  few attempts have been made to model the resource limitations.   
%However, to our best knowledge, only few architectures have been proposed to apply in resource-limted recommendation scenarios in which purchase decisions are made cautiously.

%To date, several graph models have been proposed to learn the representation of nodes(users and items) in temporal networks. CTDEN\cite{Nguyen2018ContinuousTimeDN} is state-of-art graph-based algorithm that learn static embeddings of nodes via maximization of context likelihoods of nodes. GATNE\cite{Cen2019RepresentationLF} learns multiple embeddings of different type nodes and edges, DynGem\cite{Goyal2017DynGEMDE} learns embeded time sequences of graph snapshots, which is not applicable to our continous interaction data. DGNN\cite{Ma2018DynamicGN} and DyREP\cite{Trivedi2019DyRepLR} assume that there are persistent links between nodes which is untenable in interactions netwok composed by instantaneous interactions. LDG\cite{Knyazev2019LearningTA} use two passes of node/edge embedding propagation and bilinear intensity function to estimate the interaction probability of different interaction event types between nodes in interaction graph.  

\begin{figure*}[h]
	\centering 
	\includegraphics[width=\textwidth]{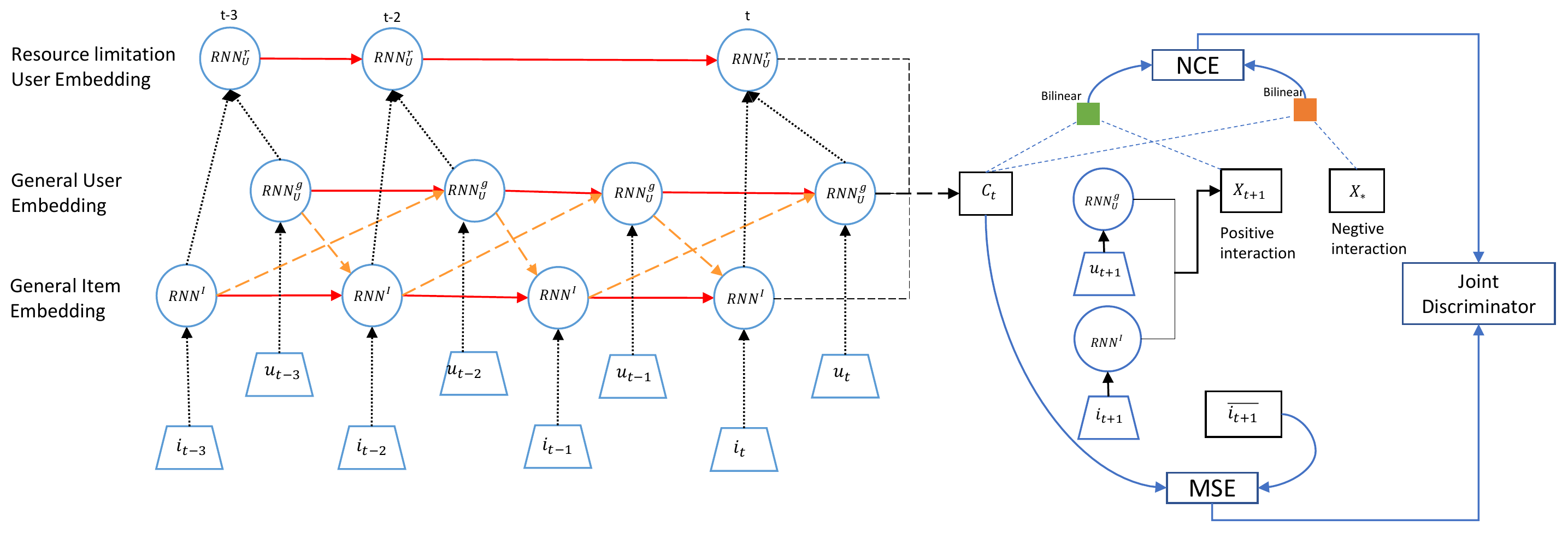}
	\caption{Illustration of our model achitecture. $\overline{i_{t+1}}$ is static one-hot embedding of $i_{t+1}$. 'NCE' are the contrastive loss function built by the mutual infomation between history context and future targets. 'MSE' are Mean Squared Error between prediction and ground truth. $RNN_U^l$, $RNN_U^g$ and $RNN_I$ are non-preference user embedding, user preference embedding and item embedding resprctively.}
	\label{figure1} 
\end{figure*}

\begin{table*}
	\caption{Table of symbols used in this paper}
	\label{tab:freq1}
	\centering
	\begin{tabular}{|c|c|}
		\toprule
		Symbol&Meaning\\ 
		\midrule
		$\mathbf{h}_u^g(t)$ and $\mathbf{h}_i(t)$&General dynamic embedding of user u and item i at time t\\
		%		$u^g(t-1)$ and $i(t-1)$&General dynamic embedding of user u and item i before time t\\
		$\bar{u}$ and $\bar{i}$&Static embedding of user u and item i\\
%		$\hat{u}^g(t)$&Projected embedding of user u at time t\\
		$\widetilde{j}(t)$&Predicted item j embedding\\
		$\mathbf{h}_u^l(t)$ &Resource-limited user dynamic embedding \\
		$c_t$&History context including user and item embedding \\
		$f_1$&Density ratio function \\
		$pn\_x_{t+1}$&positive and negtive samples of user's interactions on next timestamp\\
		\bottomrule
	\end{tabular}
\end{table*}

In information theory, mutual information is a kind of measurement for statistic dependence between two random variables $X$, $Y$ with $p(x)$, $p(y)$ representing their associated probability distributions, respectively~\cite{Yu2003FeatureWA}.  By leveraging domain knowledge, mutual information estimation has been successfully applied to a variety of problem areas, including image processing~\cite{Tian2019ContrastiveMC}~\cite{Hjelm2019LearningDR}~\cite{Hnaff2019DataEfficientIR}, video classification~\cite{Sun2019ContrastiveBT}, and natural language understanding~\cite{Oord2018RepresentationLW}. The first application of mutual information in the recommendation task is~\cite{Yu2004CollaborativeFA} which deploys mutual information to measure similarities between items' ratings of users in collaborative filtering. By treating mutual information as an item-to-item distance metric, only metric learning in item space is conducted instead of measuring the similarities between users and items based on user-item interaction. 

Different from all the aforementioned methods, in this paper, we target at the resource-limited recommendation task, and specifically construct a used car transaction dataset named `CheZhiBao'. Specifically, a RNN branch is constructed to model the status variation of users and products, and capture the influence of the limited resource for users' interests. Finally, mutual information to measure the similarities between items and users' actions for user recommendation.

%%$u \in \mathcal{U}$    $ i \in \mathcal{I} $ happening between a user and an item,
\section{Problem definition}
%\subsection{Problem definition}
The symbols in this paper are defined as follows. The sets of users and items are denoted by $\mathcal{U}$ and $\mathcal{I}$, respectively. 
 $[S_1, S_2,...S_T]$ is a time ordered sequence of temporal user-item interactions and $S_t=(u_t, i_t, \mathbf{v}_t), t \in [1,T]$. Here, $S_t$ denotes the user-item interaction between the user $u_t$ and item $i_t$ at time t. And $\mathbf{v}_t$ is an associated feature vector gathering pre-extracted descriptions of product information, including images and text description (vehicle type, registration year, miles and auction countdown ) of cars together with action types~(e.g. click). We define the resource-limited sequential recommendation problem as: given the purchase history and click history of the user $u$, we aim to predict the probability of the user-item interaction denoted as $P( u_{T+1},i_{T+1} | S_1, S_2,...S_T$) at time $T+1$ meanwhile considering the resource-limitation constraints.  Besides, for the modules in our framework in Fig.~\ref{figure1}, $RNN_{U}^{limited}$ ($RNN_{U}^{l}$ for short) means the resource-limited branch modeling the resource-limitation information, while $RNN_{U}^{general}$ ($RNN_{U}^{g}$ for short, a RNN branch modeling user interest)  and $RNN_{I}$ (a RNN branch modeling items) construct the MRRNN by learning the embeddings of  users and items respectively in a mutual way. Table \ref{tab:freq1} lists those used symbols.

\section{The Proposed Model}
In this section, we first overview the entire architecture of our proposed model, and then introduce the involved four key modules in detail, including MRRNNs, the resource-limited sub-network, the mutual information estimator and the loss function.

\subsection{Overview}
The whole structure of our proposed model is illustrated in Fig.~\ref{figure1}, and the purpose is to predict the future actions of users indicating their preferences for products. In the learning process, the pre-extracted features of users and items, together with their interactions, are passed through two branches of sub-networks for dynamic embedding, i.e. an interactive branch capturing interactive long-term user-term interaction and a resource-limited branch exploring the influence of resource variation on user interests. For the interactive branch, considering the success in previous literatures, we employ the MRRNNs to extract interactive semantics to describe the user-item interaction context. For the  resource-limited branch, another RNN is constructed to capture the variation once a transaction happens. Then, the features of both MRRNNs and the  resource-limited branch are fused, and further used to measure the similarity between user interest and the product. For tuning parameters, two types of loss functions, including a supervised loss named MSE and an unsupervised one named Personal Mutual Information, are calculated to optimize the whole network through back-propagation.

\subsection{Mutually-Recursive Recurrent Neural Networks}
\begin{figure}[h]
	\centering
	\includegraphics[width=\linewidth]{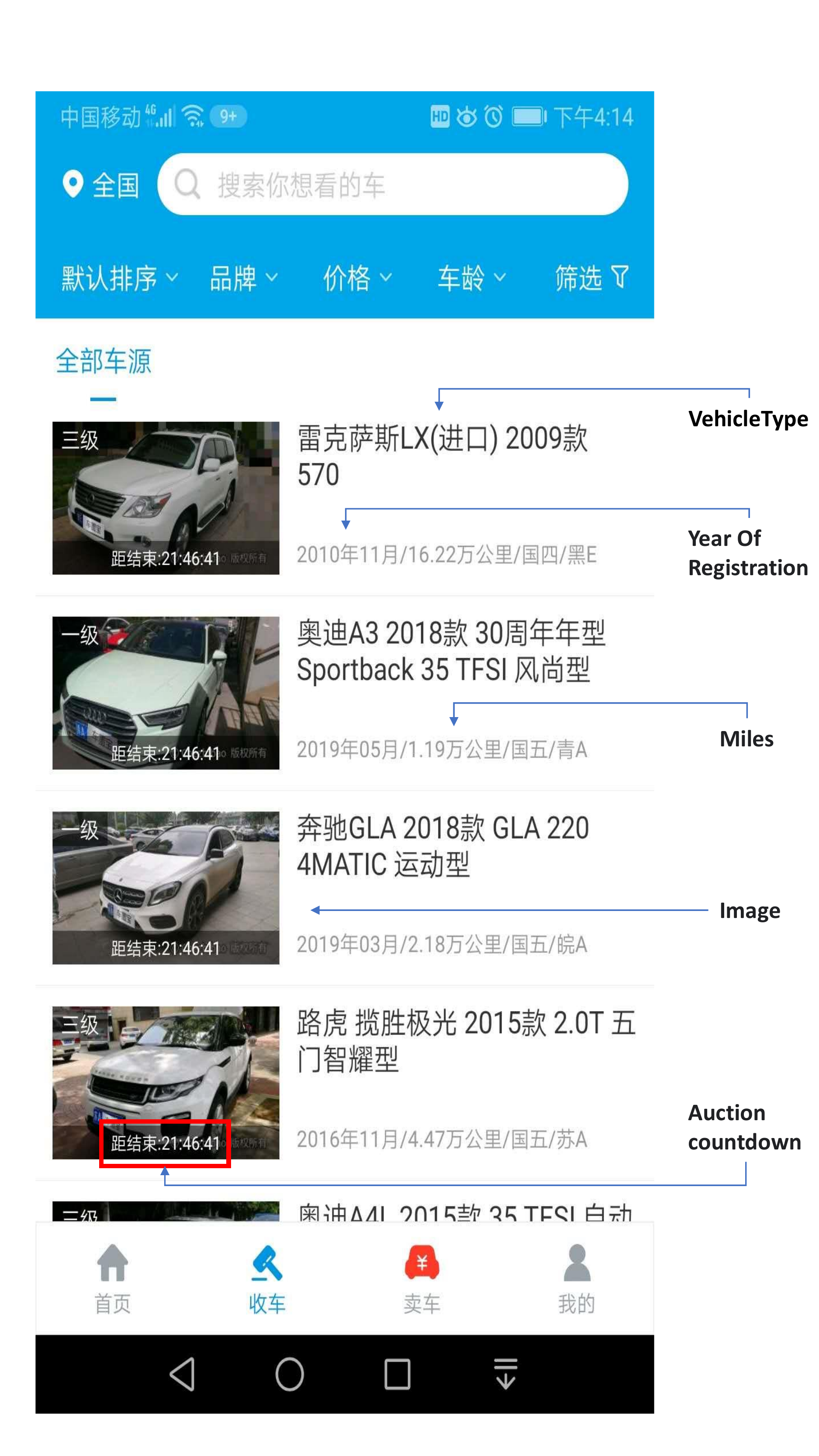}
	\caption{Item list of online used car auction App.}
	\label{figure2} 
\end{figure}

%Following JODIE\cite{Kumar2019PredictingDE}, 
To learn dynamic connection between users and items, our model leverages the time ordered interaction $ S_t=(u_t, i_t, \mathbf{v}_t) $ between user $u$ and item $i$ at time $t$ to generate two mutually recursive RNN embedding $\mathbf{h}_u(t)$ and $\mathbf{h}_i(t)$:
\begin{equation}
\mathbf{h}_u^g(t) = \sigma(\mathbf{W}_1^u\mathbf{h}_u^g(t-1) + \mathbf{W}_2^u\mathbf{h}_i(t-1)  + \mathbf{W}_3^u\Delta_u + \mathbf{W}_4^u\mathbf{v}_{t-1}),
\end{equation}
\begin{equation}
\mathbf{h}_i(t) = \sigma(\mathbf{W}_1^i\mathbf{h}_i(t-1) + \mathbf{W}_2^i\mathbf{h}_u^g(t-1)  + \mathbf{W}_3^i\Delta_i + \mathbf{W}_4^i\mathbf{v}_{t-1}),
\end{equation} where $\Delta_u$ denotes the time elapsed since user u's last interaction(with any item) and $\Delta_i$ is the time elapsed since last interaction (with any user) of item $i$. The matrices $\mathbf{W}_1^u,\mathbf{W}_2^u,\mathbf{W}_3^u,\mathbf{W}_4^u$ are the parameters of $RNN_U^g$ and $\mathbf{W}_1^i,\mathbf{W}_2^i,\mathbf{W}_3^i,\mathbf{W}_4^i$ are the parameters of  $RNN_I$. $\mathbf{v}_{t-1}$ is the interaction feature vector which involves the information of both image features and car attributes (see Fig.~\ref{figure2}). $\sigma(\cdot)$ is the sigmoid function to introduce non-linearity. 
These two embeddings learn the long-life user preference from the different kinds of interactions. The mutually recursive update of two embeddings is performed along the  timeline. 

\subsection{The Resource-Limited Branch}
In this section, a subnetwork is proposed to explore the latent effect of two resource-limited factors over user action prediction. There are two resource-limited factors, i.e. product inventory and user financial strength. The financial strength would fluctuate with the user status variation, e.g. a user's financial strength would degrade after a recent purchase of an expensive car.  And for the product inventory, it may be daily updated due to the dynamic selling intention of sellers. For instance, one specific item would become limited or out of inventory if the seller withdraws the on-sale product (e.g. a seller may turn to other online selling platforms for a better price ).

According to the analysis above, we specifically introduce a resource-limited branch to obtain user state embedding denoted as  $\mathbf{h}_u^l(t)$ once a transaction happens. As the financial strength of users may not be directly accessible in the real application, to calculate $\mathbf{h}_u^l(t)$, we represent the financial strenth with multiple related factors which implicitly or indirectly reflect the purchasing power of users. Concretely, financial strength is potentially revealed by the user purchase history (i.e. $\mathbf{h}_u^g(t-1)$), the time interval since last purchase (i.e. $\Delta_{i\_p}$), and current purchasing product (i.e. $\overline{i}_{p}$). Similarly, the product inventory is implicitily contained in the dynamic embedding (i.e. $\mathbf{h}_u^g(t)$ and $\mathbf{h}_i^g(t)$) and the decisive factors that how many cars on sales (i.e. initial inventory $I^l$) one day.
%	 are encoded in $\mathbf{h}_u^g(t)$ and $\mathbf{h}_i^g(t)$.}

The resouce-limited embedding $\mathbf{h}_u^l(t)$ is formulated as follows:
\begin{equation} 
\begin{split}
\mathbf{h}_u^l(t)=&\sigma({\mathbf{W}_1^l}{\mathbf{h}_u^l(t-1)} + \mathbf{W}_2^l\mathbf{h}_u^g(t) + \mathbf{W}_3^l\mathbf{h}_i(t) \\
&+ \mathbf{W}_4^l\overline{u}_{p} +  \mathbf{W}_5^l\overline{i}_{p}+{\mathbf{W}_6^l}\Delta_{i\_p} \\&+\mathbf{W}_7^l\mathbf{v}_{p} +\mathbf{W}_8^lI^l), 
\end{split}
\end{equation} %Including general embedding($u^g$,$i^g$) into resource-limited user embedding helps to extract some unique patterns from relatively less frequently purchase actions without leading to overfitting the purchase behavior.
where $\mathbf{W}_1^l,...,\mathbf{W}_8^l$ are parameters of $\mathbf{h}_u^{l}$. $\mathbf{v}_{p}$ is the feature vector of the purchased car. Item id $\overline{i}_{p}$ contains product price information
%, while $\mathbf{h}_i(t)$ and $\Delta_{i\_p}$ jointly model the user's personal financial strength related information implicitly.

Since some time has elapsed since user's last interaction at time t whilst make prediction at future time $t+1$, we employ a projection operation to moderately avoid user embedding's information expiration (e.g. make a prediction for one user at current now with his latest interaction happened at yestoday). Two inputs are required for the projection operation: user’s embedding at time t and the elapsed time $\Delta$.

Specifically, in Eqn.\ref{eq:4} and \ref{eq:5}, We first convert $\Delta^g$ (i.e. time elapsed since last clicked) to a time-context vector $w^g$ using a linear layer (represented by vector $\mathbf{W}_t$). The projected embedding $\mathbf{h}_u^g(t + \Delta)$ is then obtained by leverage a temporal attention vector $1 + w^g$ scaling the past user embedding $\mathbf{h}_u^g(t)$. 

\begin{equation} \label{eq:4}
w^g = \mathbf{W}_t\Delta^g,
\end{equation}
\begin{equation} \label{eq:5}
\mathbf{h}_u^g(t + \Delta) = (1+w^g)*\mathbf{h}_u^g(t).
\end{equation}
Similarly, we also do the same projection to resource-limited user embedding $\mathbf{h}_u^l(t)$:
\begin{equation} 
w^l = \mathbf{W}_t\Delta^l,
\end{equation}
\begin{equation} \label{eq:7}
\mathbf{h}_u^l(t + \Delta) = (1+w^l)*\mathbf{h}_u^l(t).
\end{equation}
The '$t + \Delta$' in Eqn.\ref{eq:5} and \ref{eq:7} means the very current timestamp (i.e. now).

In Eqn.\ref{eq:8} and Eqn.\ref{eq:12}, to model the history actions' influence on future target, we introduce an new combined user embedding $f(\mathbf{h}_u^g(t+\Delta),\mathbf{h}_u^l(t + \Delta))$ which combine (i.e. adding, concat, product, etc) the general user embedding and resource-limited user embedding. For simplicity but without loss of generality, we let $f(\mathbf{h}_u^g(t+\Delta),\mathbf{h}_u^l(t + \Delta)) = \mathbf{h}_u^g(t+\Delta)+\mathbf{h}_u^l(t + \Delta)$. The predicting target is a $one-hot$ item embedding $\bar{j}$ to be clicked or purchased. The prediction is made by a fully connected linear layer as following:
\begin{equation} \label{eq:8}
\begin{split}
\widetilde{j}(t+\Delta) =&\mathbf{W}_1^j*f(\mathbf{h}_u^g(t+\Delta),\mathbf{h}_u^l(t + \Delta))  \\
& + \mathbf{W}_2^j\bar{u} + \mathbf{W}_3^j\mathbf{h}_i(t-1) + \mathbf{W}_4^j\bar{i}(t-1),
\end{split}
\end{equation}
where $\mathbf{W}_1^j,...,\mathbf{W}_4^j$ are the parameters of the above linear layer.

\subsection{Personal Mutual Information}
%According to \cite{Oord2018RepresentationLW}, modeling  $p(future\_interaction|history\_context)$ directly may not be the most optimal for the purpose of extracting the shared high-level relation between future target and history context. 
To explore how to conduct user-item interaction behavior prediction, we propose to maximize the mutual information between the historical behavior context $c_t$ and the next-timestep action $x_{t+1}$ to learn the shared latent space information between them:

\begin{equation}
I(x_{t+1};c_t) = \sum\limits_{x_{t+1},c_t}p(x_{t+1},c_t)~log\frac{p(x_{t+1},c_t)}{p(x_{t+1})~p(c_t)}.
\end{equation}
Fig.~\ref{figure1} shows the overall architecture of our models. First, the resource-limited user embedding updates only when a transaction happens while general user/item embeddings update at every interaction (both click and transaction included). When predicting next interaction, we combine latest updated $f(\mathbf{h}_u^g(t)$, $\mathbf{h}_u^l(t)$ and item i's embedding $\mathbf{h}_i(t-1)$ as history context $c_t$. Then we inference all candidate future interaction ($\mathbf{h}_u(t+1)$,$\mathbf{h}_i(t+1)$) and make the prediction through joint ranking (Eqn.\ref{eq:20} and \ref{eq:21}). 

As discussed in \cite{Oord2018RepresentationLW}, we can model a density ratio $f_1(x_{t+1},c_t)$ which is proportional to the mutual information between $x_{t+1}$ and $c_t$:

\begin{equation}
f_1(x_{t+1},c_t) \propto \frac{p(x_{t+1}|c_t)}{p(x_{t+1})},
\end{equation}
where we define density ration $f_1$ as a simple log-bilinear model:
\begin{equation}
f_1(x_{t+1},c_t)=exp~( x_{t+1}^T\mathbf{W}_1^fc_t ).
\end{equation}

%The density ratio $f_1$ can be unnormalized(does no have to integrate to 1), and we do not use positive real score function $exp$ (equation 3 in \cite{Oord2018RepresentationLW}) which we found no helpful in model convergence  when trainning. 
%According to \cite{Tschannen2019OnMI}'s analysis about mutual information estimation in representation learning, success of mutual information estimation depends on the inductive bias in both the choice of feature extractor architectures and the parametrization of the employed MI estimators. In equation 11, we remove the exp operator for simplicity based on the effectiveness showed in experiments.

The history context $c_t$ are composed of resource-limited user embedding at the last purchase time (mapping to current time $t$), last updated general user embedding (also mapping to current time$t$) and last interacted item embedding, with letting $pos\_x_{t+1}$ be user-item interaction at future time $t+1$.

\begin{equation} \label{eq:12}
c_t = [f(\mathbf{h}_u^l(t+\Delta),\mathbf{h}_u^g(t+\Delta)),\overline{u},\mathbf{h}_i(t-1),\overline{i}(t-1)].
\end{equation}
As we can't evaluate $p(x_{t+1})$ or $p(x_{t+1}|c_t)$ directly or computationally intense, the Noise-Contrastive-Estimation\cite{Gutmann2010NoisecontrastiveEA} that based on comparing one positive sample $pos\_x_{t+1}[1]$ sampled from 'conditional interaction distribution' $p(x_{t+1}|c_t)$ and other N-1 random negtive ones $neg\_x_{t+1}[2:N]$ sampled from 'interaction distribution' $p(x_{t+1})$ is allowing us to estimate them:

\begin{equation}
pos\_x_{t+1}[1] =[\mathbf{h}_u^g(t+1),\mathbf{h}_i(t+1),\overline{i}(t+1)],
\end{equation}
\begin{equation}
neg\_x_{t+1}[2:N] =[\mathbf{h}_u^g(t+1),\mathbf{h}_i(t),\overline{i}(t)],
\end{equation}
\begin{equation}
pn\_x_{t+1} = \bigg[{pos\_x\_t+1  \atop neg\_x\_t+1}\bigg],
\end{equation}
\begin{equation}
f_1(pn\_x_{t+1},c_t) =exp~( pn\_x_{t+1}*\mathbf{W}_1^f*c_t).
\end{equation}
%\subsection{Self attention}
%as
%\subsection{Dynamic graph embedding}
%instead mutual RNN
%
%\subsection{casual inference}
%trade influence on next item 
%
%
%\subsection{few_shot learning on trade actions}

As discussed in \cite{Oord2018RepresentationLW} , the loss function is defined in terms of pair-wise similarities within a mini-batch. The performance of pair-based losses heavily rely on their capability of mining informative negative pairs. Following \cite{Wang2019CrossBatchMF}, we also desire to break the limit of mining hard negatives within a single mini-batch, so we sample $N-1$ negative pairs from all candidate item embeddings rather than (in CPC\cite{Oord2018RepresentationLW}) sample negative pairs within mini-batch which result in poorer performance. 
%The reason for our strategy can be conduct is that the number of candidate items is far less than the number of negtive data instances.  

\subsection{The loss fucntion}
We employ a joint loss of MSE(mean-squared error) Loss and NCE Loss to train three embeddings and autoregressive model.

The first part of our objective function is a MSE loss which is to minimize the sum of $L_2$ distance between the predicted item embedding and the ground truth static one avim t interaction in one batch.
\begin{equation}L_{MSE} = \sum\limits_{(u,j,t)\in{S}}|| \widetilde{j}(t+\Delta)-\bar{j} ||_2.
\end{equation}
Considering a N samples set $X = \{x_1,...X_n\}$ which contains one positive sample from $p(x_{t+1}|c_t)$ and N-1 negative samples from $p(x_{t+1})$, we define the NCE loss as:
\begin{equation}
L_{N} = -E_X\bigg[log\frac{f_k(pos\_x_{t+1},c_t)}{\Sigma_{pn\_x_j \in X} f_k(pn\_x_j,c_t)}\bigg].
\end{equation}
Optimizing $L_{N}$ is equivalent to estimating the density ratio in Eqn.(10).

By combining the above two loss items, we get fusion loss functions:
\begin{equation}
\begin{split}
Loss = \lambda_mL_{MSE} + \lambda_nL_{N} &+ \lambda_U{||\mathbf{h}_u^g(t)-\mathbf{h}_u^g(t-1)||}_2\\&+\lambda_I{||\mathbf{h}_i(t)-\mathbf{h}_i(t-1)||}_2,
\end{split}
\end{equation}
where we add term $\lambda_n$ to control the influence of the mutual information part on the joint training and term $\lambda_m$ to control the influence of MLP projection. 
The last two regularization terms we add ensure the general interesting "slow drift" phenomena\cite{Wang2019CrossBatchMF} that the embedding of instance~(user and item) actually drifts at a relatively slow rate when the model tend to convergence. To verify the "slow drift" phenomena, we remove two regular terms and find that the performance of our fusion model descends a lot.

During training for the different datasets, the fusion strategy needs to be carefully designed. Each dataset has its own specific business characteristics. 
%Whether restricted user embedding $\hat{u}^r(t+\Delta)$ in $\widetilde{j}(t + \Delta)$ ( equation 8) or $c_t$ (equation 12) depends on the characteristics of the data.

We treat the fusion loss as a joint trainning multi-task object, and the RNN embedding can be substitued by LSTM or enhanced by self-attention. We leave these improvements in future work.

\subsection{Differences between IMN-Net and JODIE}
 Our research differs from the JODIE model in three general perspectives. First, our model mainly focuses on the resource limitation fators' representation learning on bulk commodities e-commerce websites while JODIE mainly studies the general user behavior pattern on conventional non-e-commerce websites like social and music websites. 
Then, there are different perspectives on model design. JODIE  concentrates on the long-short term preference model design paradigm. Our model introduces two different user states, one of which is a resource-limited user embedding while the other is embedding of general user preference. Moreover, we design a contrastive loss to learn  latent connections between history purchase actions influent future interactions by maximizing mutual information of them.  
Finally, there are different hypotheses between JODIE and our proposed model. JODIE assumes that the user state is slowly transforming and stable to some extend. The inductive bias behind our model also includes that the user's internal state potentially contains latent resource limitation elements that have a special distribution when in bulk commodities recommendation system.

\section{Experiments}
In this section, we conduct experiments on two commercial datasets respectively, and compare our model IMN-Net with five well known baselines of sequential recommendation.
%\begin{itemize}
%	\item \textbf{RQ1:} How does our proposed IMN-Net perform as compared with state-of-the-art recommendation systems that are designed for representation learning in sequential recommendation?
%	\item \textbf{RQ2:} Why do we design this model fused of supervised and mutual information estimator like this way?
%	\item \textbf{RQ3:} What are the effectiveness of mutual information estimator and resource-limited user embedding?
%\end{itemize}
\begin{table*}
	\caption{Dataset statistics }
	\label{tab:freq2}
	\centering
	\begin{tabular}{ccccccccc}
		\toprule
		\midrule
		Dataset&\#users&\#items&\#actions &$\frac{actions}{users}$ &$\frac{actions}{items}$&time span&$purchase \atop ratio$&price (CNY)\\
		\midrule
		CheZhiBao 2019&10,718&40,820&647,664&378&13&2019.5 - 2019.10&2.42\%&150k\\ 
		Tmall/Koubei 2016&1,160,135&2,353,208&44,528,128&46&19&2015.7 - 2015.11&24.5\%&1k\\ 
		\bottomrule
	\end{tabular}
\end{table*}
\begin{table*}
	\caption{Future Interaction Prediction}
	\label{tab:freq3}
	\centering
	\begin{tabular}{ccccccccccc}
		\toprule
		\midrule
		Datasets&Metrics&FPMC&SASRec&GRU4Rec&NextItNet&JODIE&IMN\_MSE&IMN\_NCE&IMN\_Cos&IMN-Net\\ 
		\midrule
		&Recall@10&0.0554&0.0414&0.0320&0.0228&0.0659&0.0642&0.0480&0.0527&$\textbf{0.0856}$\\
		CZB&NDCG@10&0.0274&0.0264&0.0159&0.0112&0.0527&0.0518&0.0355&0.0410&$\textbf{0.0660}$\\
		&Recall@20&0.0933&0.0640&0.0589&0.0413&0.0933&0.0950&0.0783&0.0854&$\textbf{0.1281}$\\ 
		&NDCG@20&0.0432&0.0292&0.0226&0.0158&0.0627&0.0630&0.0463&0.0529&$\textbf{0.0815}$\\
		\midrule
		&Recall@10&0.4012&$\textbf{0.5108}$&0.3211&0.3917&0.4014&0.4107&0.3429&0.3602&0.4296\\ 
		Tmall&NDCG@10&0.3089&0.4862&0.2484&0.3305&0.5089&0.5169&0.3939&0.4388&$\textbf{0.5393}$\\ 
		&Recall@20&0.4503&$\textbf{0.5256}$&0.3675&0.4256&0.4148&0.4247&0.3822&0.4077&0.4562\\ 
		&NDCG@20&0.3213&0.4900&0.2600&0.3391&0.5138&0.5221&0.4092&0.4214&$\textbf{0.5490}$\\ 
		\bottomrule
	\end{tabular}
\end{table*}

\subsection{Dataset Desciption}
The first dataset 'CheZhiBao\footnote{https://www.chezhibao.com/} 2019' (CZB for short) is built from the biggest online used car auction website in China. This website is a used car auction selling website for the whole of China. Each car will be sold to the buyer within one week. On this platform, quantity and categories of commodities on the good shelves change every day. The average price of commodities is more than 10,000 dollars. Most users in CheZhiBao dataset are used car sales enterprises, who come to the platform every day and click products to capture market dynamics as well as find new products. This yields a number of click/transaction sequences. For non-active users, they may be individual customers with rather limited activity. Considering this huge user activity differentiation, the influence of non-active users can be neglected in the dataset construction without degrading the recommendation performance. For each user, we adopt all of their interaction data without filtering. In this dataset, there are three kinds of user actions: bid, win, trade after entering (clicked) a car auction page. We treat purchase actions as 'purchase' interactions while the other two kinds of actions as 'click'. We obtain the image feature using vgg16 features while the structure feature using id embedding. 

The second data set is extracted from Tmall, the largest B2C platform in China. It is a dataset obtained from Tmall/Koubei IJCAI16 Contest\footnote{https://tianchi.aliyun.com/dataset/dataDetail?dataId=53}(Tmall for short). This data set comes from a discount coupon APP, from which users mainly order for consumption discount. There are two behaviors in the dataset: click and purchase. 
%Because of the low average price of goods purchased by users in the data set, we randomly sample 3000 users' actions according to our problem definition, and expected the selected users to cover a certain proportion of high consumption users. 
%
%To preprocess both of two datasets, our model adopts all the user interactions without filtering items set. 
And we seek to make our model be a general framework for the e-commercial purchase-aware sequential recommendation, so we adopt only two main dimensions, 'user' and 'item', of the dataset. The statistics of the two datasets are summarized in Table \ref{tab:freq2}.
\subsection{Experimental settings}
For our model and all baselines, the datasets are split by time order. We use first 80\% interactions as train set, next 10\% interactions as the validation set, and last 10\% as the test set. We set the dimensions of the dynamic embeddings of all models to 128. We use Adam to minimize the loss, with learning rate of $10^{-3}$ and mini-batches of 256 records. All models run 50 epochs and report test results of them corresponding to the best performing validation set.
Following the JODIE setting, our model IMN\_* uses t-batch\cite{Kumar2019PredictingDE} for training data mini-batch.   

%For all datasets, IMN\_MSE, IMN-Net, IMN\_NCE, JODIE, NextItNet, SASRec, and GRU4Rec use the learning rate of 0.001. FPMC uses the learning rate of 0.01.
\subsection{Evaluation Protocols}
We evaluate all models with two popular ranking-based metrics: Recall@K and Normalized Discounted Cumulative Gain (NDCG). Recall@K measures the proportion of the top-K recommended items that are in the real candidate ones, We adopt K = \{10,20\}. NDCG is a ranking metric. In the context of sequential recommendation, it is formulated as NDCG = $\frac{1}{log_2(1+rank_{pos})}$, where $rank_{pos}$ is rank of correctly predicted items.
\subsection{Baselines}
\begin{itemize}
	\item \textbf{SASRec}\cite{Kang2018SelfAttentiveSR} is a self-attention based sequential model, and it can consider consumed items for next item recommendation.
	\item \textbf{NextItNet}\cite{Yuan2019ASC} applies 1D CNNs with dilated convolution filters and residual blocks to model sequential recommendation.
	\item \textbf{GRU4Rec}\cite{Hidasi2015SessionbasedRW} applies GRU to model user click sequences for the session-based recommendation.
	\item \textbf{FPMC}\cite{Rendle2010FactorizingPM} fuses matrix factorization and first-order Markov Chains to capture long-term preferences and short-term item-item transitions, respectively, for next item recommendation.
	\item \textbf{JODIE}\cite{Kumar2019PredictingDE} introduce mutually recursive RNN to model general user and item embedding for sequential recommendation.
\end{itemize}

\subsection{IMN-Net's Parameters and Results}
\textbf{Parameters}: To evaluate the performance of our proposed method, we compare it with five state-of-the-art methods.
The fusion loss in Eqn.(19) is adopted to optimize all the embeddings and the network parameters. The hyperparameter N in Eqn.(14) is searched in $\{32,64,128,256,320\}$ and setting N to 128 resulting in the best performance. 
As can be concluded from section2.3 in CPC\cite{Oord2018RepresentationLW}, the mutual information between the context $c_t$ and target $x_{t+k}$ becomes tighter as N becomes larger. 
During tuning, we found that the best N is proportional to the size of the dataset. The other two hyperparameters $\lambda_m$ and $\lambda_n$ can be tuned according to the characteristics of datasets. 
%For simplification, both of them are set as 0.5 for two datasets.

Since the huge number of users and items will lead to large number of parameters of Bilinear parameter $W_1$ in Eqn.(11). So two id compression embeddings for user and item static one-hot encodings are added. These two compression embeddings' input and output dimensions are $[N_{user},128]$ and $[N_{item},128]$ respectively where $N_{user}$ and $N_{item}$ are the total numbers of users and items.

Our proposed model is a multi-task joint learning framework, the prediction is the item with highest-ranking position. During validation and testing stage, we apply the fused version of two metric distance (MSE and mutual information) as a \textbf{joint discriminator} to make composite ranking in all candidates of items to choose the highest ranking item:
\begin{equation} \label{eq:20}
\begin{split}
fusion\_distance(i) = &\alpha_m|| \widetilde{j}(t+\Delta)-\bar{j}_i ||_{i\in[1,...N_{item}]}  \\ &+\alpha_nf_1(x_i,c_t)_{i\in[1,...N_{item}]},
\end{split}
\end{equation}
\begin{equation} \label{eq:21}
i_{t+1} = \mathop{\arg\max}_{i\in[1,...N_{item}]}\{fusion\_distance(i)\},
%i_{t+1} =  i_{max\{fusion\_distance_i\}}
\end{equation}
 where $i_{t+1}$ is the predicted item id at timestamp.
 
The hyperparameters $\alpha_m$ and $\alpha_n$ are applied greedy searches on both two datasets respectively. The best values of these two hyperparameters are $\{0.75,0.25\}$ and $\{0.9,0.1\}$ for Tmall and CZB datasets respectively. 

\textbf{Results}: The results in Table 3 illustrate that our fusion model IMN-Net outperforms all baselines except RECALL@k metrics of model SASRec and \{Recall@10,NDCG@10\} of JODIE. We analyze that the advantage in Recall metrics of SASRec over IMN\_* is mainly due to the characteristics of datasets since the Tmall/Koubei dataset mainly contains users’ daily repeated consumption, such as catering, clothing consumption, etc. This daily consumption's scale is not enough to have a fundamental impact on users' resource-limited factors, which is exactly the focus of our research problem definition. Self-attention (e.g. SASRec) is a good method for modeling repeated short-term behavior patterns. But for the case of interaction sequence without repetitive patterns, such as bulk commodities trading, its performance cannot catch up with our proposed model.

Comparing evaluation results of model IMN\_NCE/IMN\_MSE and model IMN-Net in Table 3, we find that the mutual information constraint and resource-limited branch significantly helps to improve on NDCG@k metrics which means that mutual information estimation has advantages on improving recommending accuracy. Especially on CheZhiBao dataset, IMN-Net is ahead of all baselines.

\section{Ablation Study}
To further show the effectiveness of our proposed resource limitation branch and mutual information constraint for representation learning, we conduct the following three experiments:

\textbf{Efficacy of resource limited branch}: In the variant model IMN-Net$^*$, the resource limitation branch $\mathbf{h}_u^l(t + \Delta)$ is removed from the combined user embedding (used in Eqn.(11,15)) while other parts of IMN-Net$^*$ are consistent with IMN-Net. The experiment's settings for this variant are the same as IMN-Net. The results in Table 6 show that the IMN-Net$^*$'s performance degrade compared with IMN-Net, on both datasets, which implies that the resource limitation branch plays a crucial role in IMN-Net. The effectiveness of resource limited branch is also evidenced that, observed in Table 3, IMN\_MSE can beat JODIE on all metrics significantly on Tmall dataset while also improve JODIE on Recall@20 and NDCG@20 on the CZB dataset.

\textbf{Efficacy of NCE Loss}: To show the effect of NCE loss, in the variant IMN\_MSE, the NCE loss is abandoned from IMN-Net while both of the MSE loss and two regularization terms are preserved:
\begin{equation}
\begin{split}
Loss = \widetilde{L}_{MSE} = &\sum\limits_{(u,j,t)\in{S}}|| \widetilde{j}(t+\Delta)-\bar{j} ||_2 			\\&+\lambda_U{||\mathbf{h}_u^g(t)-\mathbf{h}_u^g(t-1)||}_2\\&+\lambda_I{||\mathbf{h}_i(t)-\mathbf{h}_i(t-1)||}_2
\end{split}
\end{equation}

We employ $f(\mathbf{h}_u^g(t+\Delta),\mathbf{h}_u^l(t + \Delta))$ in Eqn.(11) to model the resource limitation while the other parts of the model are the same as IMN-Net. We compare the performance of IMN\_MSE with IMN-Net in Table 3, it can be seen that, on both datasets, IMN\_MSE's performance drops a lot compared with IMN-Net which shows NCE loss's effectiveness in performance improvement.
\begin{table}
	\caption{The visualization of groud truth and predictions (limited financial strength case)}
	\label{tab:freq4}
	\centering
	\begin{tabular}{|c|c|c|}
		\midrule
		date&GT&Prediction\\ 
		\midrule
		2019-1-1&Lexus NX (imported) Lexus&Audi A6L Audi\\
		2019-1-7&BMW 7 Series (imported)&BMW 5 Series\\
		2019-1-11&Cadillac XTS Cadillac&Audi Q5l, Audi\\
		2019-1-16&Sparrow Honda&\textbf{Maiteng Volkswagen}\\
		2019-1-18&A8L (imported) Audi&\textbf{BMW X1 BMW}\\
		2019-1-18&Ghibli (import) Maserati&\textbf{Audi A3 Audi}\\
		2019-1-18&\textbf{Kia K3 Kia}&BMW M2 BMW\\
		2019-1-29&\textbf{BJ 212 made by BAIC}&\textbf{Passat Volkswagen}\\
		2019-2-28&BMW X5 (imported) BMW&Touareg Volkswagen\\
		2019-3-7&\textbf{BMW 1 Series BMW}&BMW X6 BMW\\
		2019-4-11&BMW X5 (imported) BMW&Benz GLC class Benz\\
		2019-5-7&Eulogizing CDX&BMW 4 series BMW\\
		2019-5-7&Mercedes Benz C-class Benz&Levante Maserati\\
		2019-5-7&Ruizhi Toyota&Benz GLE Benz\\
		2019-5-22&Audi A5 (imported) Audi&BMW 6 Series GT BMW\\
		2019-6-5&Chuangku Chevy&\textbf{BMW X1 BMW}\\
		2019-6-12&\textbf{Accord Honda}&BMW 3 series BMW\\
		2019-6-19&Mercedes Benz C-class Benz&\textbf{Tiguan L Volkswagen}\\
		2019-6-22&\textbf{Maiteng Volkswagen}&E-class Mercedes Benz\\
		2019-6-26&Mercedes Benz C-class Benz&\textbf{Golf Volkswagen}\\
		2019-6-26&\textbf{Lacrosse Buick}&\textbf{Audi A4L Audi}\\
		2019-7-18&\textbf{BYD F3 BYD}&\textbf{Sidy Honda}\\
		
		\bottomrule
	\end{tabular}
\end{table}

\begin{table}
	\caption{The visualization of groud truth and predictions (inventory constraint case)}
	\label{tab:freq6}
	\centering
	\begin{tabular}{|c|c|c|}
		\midrule
		date&GT&Prediction\\ 
		\midrule
		2019-1-4&Roewe RX8&Pentium X40 Pentium\\
		2019-1-8&Chang'an cs55 Chang'an&Yuan new energy BYD\\
		2019-1-10&Qichen m50v Qichen&BJ 212 made by BAIC\\
		2019-1-18&Yidong DT Chang'an&Fengjun 6 great wall\\
		2019-1-18&Yidong DT Chang'an&Kia K4 Kia\\
		2019-1-21&Roewe rx5&KX cross Kia\\
		2019-2-20&Don BYD&Fit Honda\\
		2019-3-20&Roewe rx5&Havel H5 Havel\\
		2019-3-26&Vision X3 Geely Automobile&Jinke Nissan\\
		2019-4-1&Popular SX6, Dongfeng popular&Fengjun 5 Great Wall\\
		2019-4-2&\textbf{Tiguan L Volkswagen}&\textbf{Tiguan L Volkswagen}\\
		2019-4-25&\textbf{BMW 1 Series BMW}&Ruihu 3x Chery\\
		2019-4-28&Southeast DX7 Southeast&C30 Great Wall\\
		2019-5-30&\textbf{Toyota CR-V Toyota}&\textbf{Toyota CR-V Toyota}\\
		2019-6-11&Anconway Buick&Otto Suzuki\\
		2019-6-18&Ruihu 8 Chery&Song Pro BYD\\
		2019-6-19&Song new energy BYD&Song max BYD\\
		2019-6-21&Song new energy BYD&Don BYD\\
		2019-6-21&\textbf{Encora, Buick}&\textbf{BMW 1 Series BMW}\\
		2019-6-27&Leading the public&Yuan new energy BYD\\
		2019-9-11&Tang new energy BYD&Song new energy BYD\\
		2019-9-17&Song Pro BYD&Song Pro BYD\\
		
		\bottomrule
	\end{tabular}
\end{table}

%In equation 3, we found that maybe this user non-preference embedding module do not include indicator that which user is the comsumer of the $\bar{i}_{trade}$. so we add static user one-hot coding to equation 3 like this:

%\begin{equation}
%	u^o(t)=\sigma({W_1}{u^o(t^{-})} +  W_3\overline{u}_{trade} + W_2\overline{i}_{trade}+ {W_4}\Delta_i)
%\end{equation}

To further show the effectiveness of the mutual information constraints, in the variant IMN\_NCE, the whole MSE loss is discarded. The model was trained following unsupervised style by using NCE loss alone :
\begin{equation}
%\begin{split}
Loss = L_N +\lambda_U{||\mathbf{h}_u(t)-\mathbf{h}_u(t-1)||}_2 +\lambda_I{||\mathbf{h}_i(t)-\mathbf{h}_i(t-1)||}_2
% \widetilde{L}_{N} = L_N 
%&+\lambda_U{||u(t)-u(t-1)||}_2\\&+\lambda_I{||j(t)-j(t-1)||}_2
%\end{split}
\end{equation}

It is revealed in Table 3 that mutual information constraint optimization can achieve competitive performance, on CZB dataset, compared with supervised baselines: FPMC, SASRec, GRU4Rec, NextItNet. But in the Tmall dataset, the mutual information constraint shows relatively poor performance which may with the reason that on a daily consuming e-commerce platform, the latent connections between historical purchase actions and future user-item interactions (i.e. resource limitation impact) is hard to learn since some users' actions of low customer unit price may have little impact on resource inventory or capital strength. The other reason for poor performance may due to the release rhythm of sellers' discount coupons budget which depends on the sellers' business strategy. During this experiment, we use the following  distance metric for evaluation and testing:
\begin{equation}
MI\_distance(i) = f_1(x_i,c_t)_{i\in[1,...N_{item}]}
\end{equation}\begin{equation}
i_{t+1} = \mathop{\arg\max}_{i\in[1,...N_{item}]}\{MI\_distance(i)\}
%i_{t+1} =  i_{max\{MI\_distance_i\}}
\end{equation}

As discussed in \cite{Hnaff2019DataEfficientIR,Tschannen2019OnMI}, most of the researches about mutual information's application are focusing on the areas of image, audio, etc. Despite some works\cite{Hjelm2019LearningDR}\cite{Tian2019ContrastiveMC}\cite{Oord2018RepresentationLW}\cite{Hnaff2019DataEfficientIR} successfully obtain promising results via mutual information maximization, while other research\cite{Tschannen2019OnMI} shows that maximizing tighter bounds on mutual information may lead to worse representation learning. This evidence shows that we could not only depend on the mutual information estimation framework to make representation learning of dynamic embedding but also need to add some other auxiliary submodules which may better to be learned by additional unsupervised/supervised loss. Since these considerings, we design the joint discriminative framework and conduct above three experiments to prove the effectiveness of our design. 

In order to explain the reason why we take mutual information maximization to learn the connections between the user's historical actions and future ones, we choose another cosine distance loss, which is commonly used in metric learning, to construct a variant IMN\_Cos of IMN-Net. We compared the performance between IMN\_Cos and IMN-Net in Table 3, it can be seen that cosine distance version variant's performance is even worse than that of IMN\_MSE, let alone IMN-Net. 

%For simplicity, we mainly focus on the improvement of fused version model IMN-Net over our main baseline JODIE\cite{Kumar2019PredictingDE}. As showing in Table 4, the learning framework integrating mutual information constraint improves all metrics on both datasets. The metrics of the column 'IMN-Net$^*$' show that resource-limited user embedding plays a very important role.

%It is proved by the above two experiments that our approach is to not only solve the problem of bulk commodity (i.e. CheZhiBao) recommendation, but it also improves performance on daily consuming e-commerce (i.e. Tmall/Koubei) datasets.

\textbf{Analysis of Resource-limited Factors Learning}:

In order to prove the rationality of our motivation that the inventory constraint or limited financial strength have a decisive impact on the future behavior of users, we present some ground truth and predictions to reveal the underlying laws of data and effectiveness of the model. Column GT and Predictions are user actions and recommendation results respectively.

We randomly sample two user's interactions from CheZhiBao dataset, the ground truth, and recommendation results are shown in Table 4 and Table5. Table 4 is the visualization of the limited financial strength case. We can see that this user has a shortage of funds in January and June. He chose to buy some relatively cheap cars (annotated  bold in the GT column). The proposed model has caught the user's financial constraints in advance and recommended alternative cars (annotated  bold in the prediction column). Table 5 is the  visualization of inventory constraint case. User in Table 5 mainly purchases low-end SUV vehicles, but sometimes when the inventory of the arget model is not available, the user also purchases other candidate vehicles recommended by our model.

\begin{table}
	\caption{Effectiveness of mutual information estimator and resource-limited branch}
	\label{tab:freq5}
	\centering
	\begin{tabular}{c|c|c|c|c}
		\toprule
		\midrule
		Datasets&Metrics&JODIE&IMN-Net&IMN-Net$^*$\\ 
		\midrule
		&Recall@10&0.0659&0.0856&0.0810\\
		CZB&NDCG@10&0.0527&0.0660&0.0618\\
		&Recall@20&0.0933&0.1281&0.1227\\ 
		&NDCG@20&0.0627&0.0815&0.0769\\
		\midrule
		&Recall@10&0.4014&0.4296&0.4254\\ 
		Tmall&NDCG@10&0.5089&0.5393&0.5364\\ 
		&Recall@20&0.4148&0.4562&0.4530\\ 
		&NDCG@20&0.5138&0.5490&0.5464\\ 
		\bottomrule
	\end{tabular}
\end{table}

\section{CONCLUSION}
In this paper, we argue that the user's history purchase context has an important influence on all kinds of future interaction prediction which has been ignored in the previous literature. Our proposed model uses two user state embeddings to capture user's general preference and resource-limited one respectively. Also, we use mutual information estimator to help to improve the representation learning. We show that our model achieves a competitive result in the sequential recommendation of bulk commodities compared with the state-of-the-art. In future work, we will study how to learn representations that separate the explanatory latent factors behind the data (e.g. merchant inventory, user inventory, capital strength, display strategy) rather than only two rough branches in IMN-Net.

%
%
%\begin{figure}[h]
%  \centering
%  \includegraphics[width=\linewidth]{sample-franklin}
%  \caption{1907 Franklin Model D roadster. Photograph by Harris \&
%    Ewing, Inc. [Public domain], via Wikimedia
%    Commons. (\url{https://goo.gl/VLCRBB}).}
%  \Description{The 1907 Franklin Model D roadster.}
%\end{figure}

\ifCLASSOPTIONcompsoc
\section*{Acknowledgments}
\else
\section*{Acknowledgment}

\fi
The authors would like to thank the anonymous reviewers for their constructive comments and the School of Computer Science and Engineering for financial support.
\bibliographystyle{IEEEtran}
\bibliography{reference.bib}

\end{document}